\documentclass{elsart}
\usepackage{epsfig}

\journal{Nuclear Instruments and Methods A}

\begin{document}

\begin{frontmatter}

\title{Drift chamber readout system of the DIRAC experiment}

\author{L.Afanasyev\corauthref{cor1}},
\author{V.Karpukhin},

\address{Joint Institute for Nuclear Research, Dubna, Russia}

\corauth[cor1]{Corresponding author:
E-mail: Leonid.Afanasev@cern.ch,\\
Phone: +7 09621 62539, Fax: +7 09621 66666\\
Mail address: Joint Institute for Nuclear Research, \\
Dubna, Moscow Region, 141980 Russia\\
Also at: CERN EP Division, CH-1211 Geneva 23, Switzerland}

\begin{abstract}
  A drift chamber readout system of the DIRAC experiment at CERN is
  presented.  The system is intended to read out the signals from
  planar chambers operating in a high current mode. The sense wire
  signals are digitized in the 16-channel time-to-digital converter
  boards which are plugged in the signal plane connectors. This design
  results in a reduced number of modules, a small number of cables and
  high noise immunity.  The system has been successfully operating in
  the experiment since 1999
\end{abstract}

\end{frontmatter}

\section{Introduction} \label{sec:intr}

The two-arm spectrometer of the DIRAC experiment \cite{proposal} at CERN
includes a set of planar drift chambers with about 2000 signal
wires in the $X$, $Y$ and $W$-planes. Hit wire signals are digitized in
on-chamber mounted time-to-digital converters (TDCs). Numbers of hit
wires in $X$-planes are processed in the trigger system of the setup to
find the tracks in both arms of the spectrometer. The tracks found are
compared with respect to the relative momentum of pion pairs \cite{trig}
and the accepted events are collected during beam spill in the VME
buffer memories.

A feature of these drift chambers is the high current mode of
operation \cite{dc} i.e. the signals from sense wires typically have an 
amplitude of 1~mA and a width about 20~ns at half-maximum.  

The on-chamber mounted TDC board is the main unit of the readout
system.  It combines 16 input signal discriminators, the 16-channel
TDC itself and a data buffer. The boards are immediately plugged in
the signal plane connectors. Up to 8 boards can be read out via a
common bus. In comparison with the standard configuration, in which
the discriminators and the TDCs are separate modules and chambers
operate in the proportional mode, the described system has the
following advantages:

\begin{itemize}
\item The number of modules in the system is reduced; 
  
\item The number of cables is reduced. There are no external connections
  between the discriminators and the TDCs;
 
\item There are no the time measurement distortions due to crosstalks
  in a long flat cable, which usually connects the discriminators and
  the TDCs in the standard configuration. Note that crosstalks between the
  adjacent channels in a flat cable 30 m long cause a pulse
  jitter about 1.5~ns;
  
\item Due to the high threshold, 0.1~mA, of input discriminators
  practically all external pickup noise is rejected.
\end{itemize}
        
\section{Readout system} \label{sec:read}

\begin{figure}[htbp]
\begin{center}
\epsfig{file=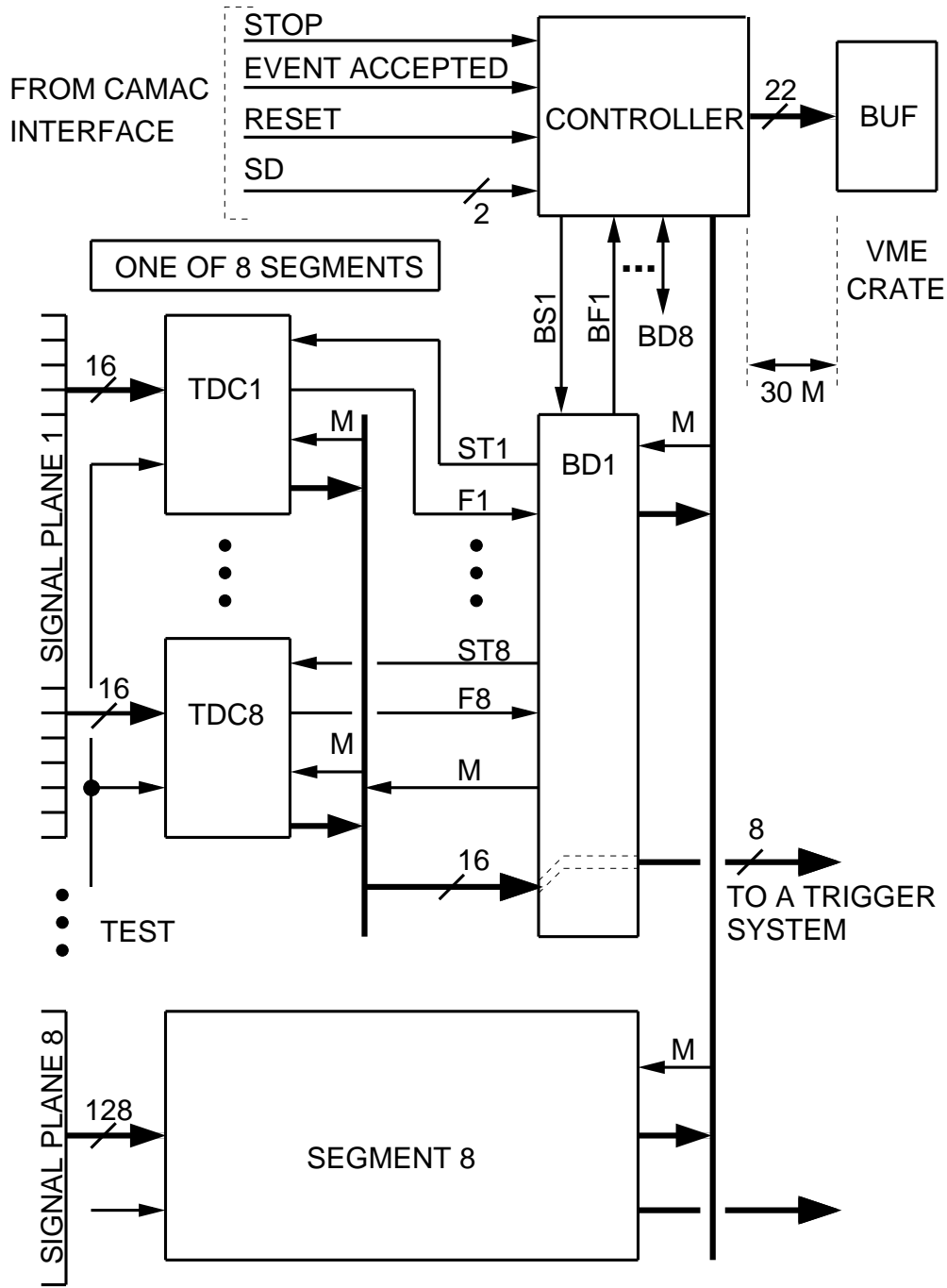,width=0.95\textwidth}
\end{center}
\caption{Block diagram of the readout system.}
\label{fig1}
\end{figure}

The readout system consists of 8 segments, which handle the hit wire
signals in parallel (Fig.\ref{fig1}).  After RESET the TDC boards
continuously accept the hits until a common STOP signal arrives at the
readout system.  It causes acquisition to stop and 
processing of the recorded hits to start. The hits found inside a TDC time
window are loaded into the on-board buffer memory. After all TDCs in
the segment have completed the data processing, the selected hit data
are transferred from the buffers to the bus driver (BD) and, in
parallel, to the trigger system (hit wire numbers only). The TDC
boards are read out in a geographical addressing mode, empty buffer
memories are skipped.  Due to a pipeline structure of the TDC boards
the recorded data are processed and the selected ones are read out at
the rate of 100~ns per hit.

Depending on state of the control bus M the individual lines ST
are used to fan out the common STOP signal or to read out the TDC buffer.
The flag F indicates the board status. Similarly the BS and BF lines
control the bus driver. With respect to the common STOP the hits
selected in a segment will be available for processing in the trigger
system at the access time:

$$
\mathrm{T_a = T_0 + (N_{rm} + N_{ss} + 3)\cdot 100\;ns}\;,
$$

where $\mathrm{T_0} = 700$~ns is the constant delay, $\mathrm{N_{rm}}$
is the maximum number of hits recorded in one of all TDC boards,
$\mathrm{N_{ss}}$ is the number of hits selected in all TDCs of the
segment and 3 is the number of ``pumping'' clocks.

To minimize the access time each $X$-plane is connected to only one
segment of the readout system.

The selected hits are stored in the bus drivers until the readout
controller receives the EVENT ACCEPTED or RESET signal from the
trigger system. An accepted event is numbered and transferred into the
VME buffer memory (BUF) at the rate of 100~ns/hit. The RESET discards
all temporary recorded data. During the beam spill the readout system
is hardware controlled. It requires only three external control
signals: STOP, EVENT ACCEPTED and RESET. During the interval between
the spills the VME processor reads out all VME buffers \cite{daq}.

A two-wire bus (SD) is used to set the TDCs time window.

The bus drivers and the controller are placed near the drift chambers in a
NIM crate, which is also used to power the readout system. Due to the
setup layout three such systems are used to read out all drift chambers.

\section{TDC board}\label{sec:tdc}

The 16-channel TDC board discriminates, digitizes, selects and buffers
the hit wire signals. The last 16 hits in each channel can be
recorded. The TDC operates in the COMMON STOP mode therefore no
additional delay in the recording channel is required.

\begin{figure}[htbp]
\begin{center}
\epsfig{file=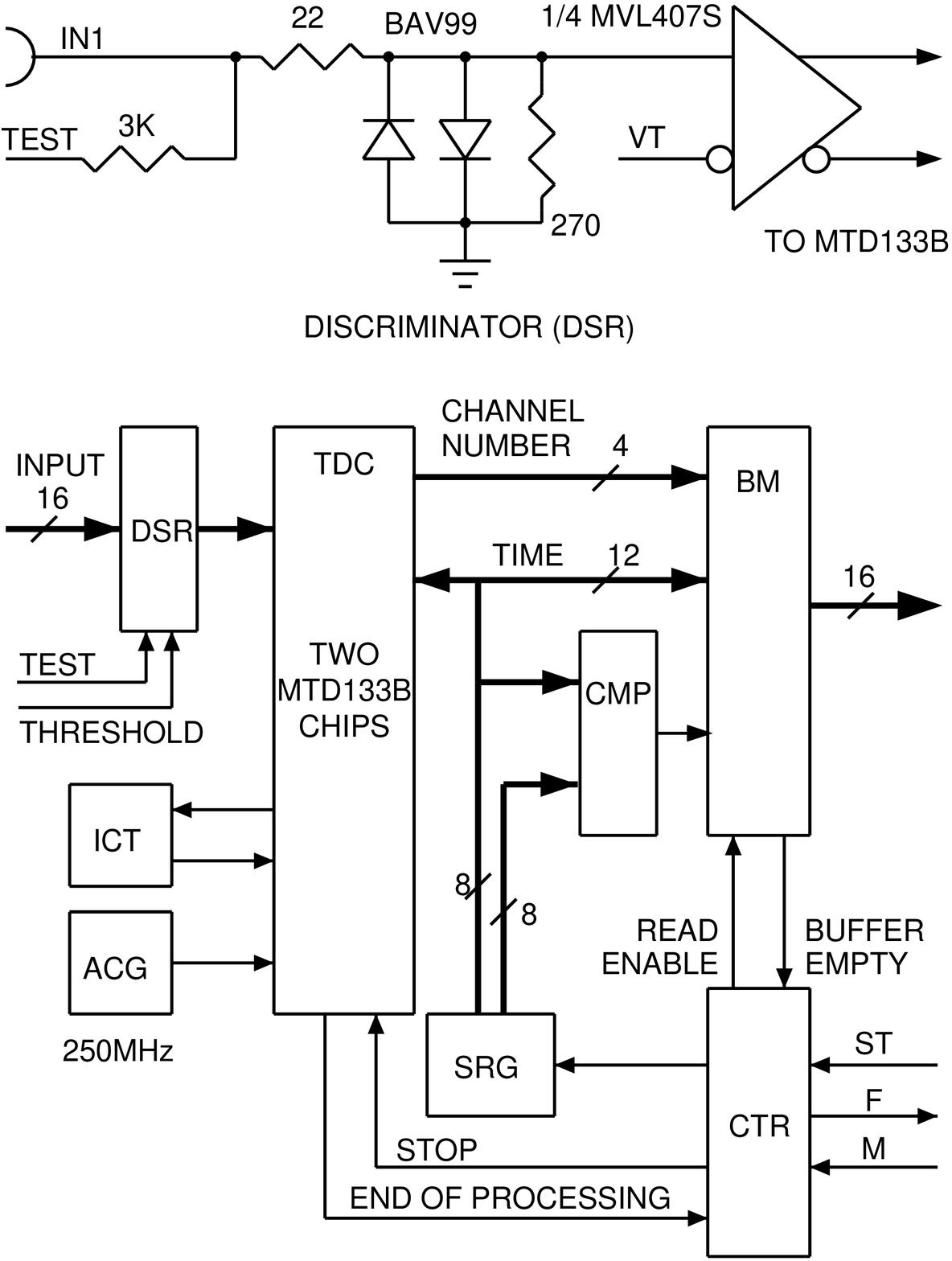,width=0.9\textwidth}
\end{center}
\caption{Block diagram of the TDC board. DSR --- discriminator, TDC ---
  time-to-digital converter, CMP --- magnitude comparator, BM ---
  buffer memory, CTR --- control unit, SRG --- shift registers, ICT
  --- interpolator control, ACG --- acquisition clock generator, VT
  --- threshold voltage.}
\label{fig2}
\end{figure}

The TDC board (Fig.\ref{fig2}) consists of an input discriminators
(DSR), a time-to-digital converter (TDC), a magnitude comparator
(CMP), a buffer memory (BM), and a control unit (CTR). 

Two 8-bit shift registers (SRG) are used to set the maximum time
measurement range and the offset value, i.e. the far and near time
thresholds, respectively.  The TDC time window is defined as
difference between these thresholds.

The input discriminator consists of a protection scheme (BAV99) and an
ultrafast voltage comparator MVL407S \cite{mvl} (one chip per four
channels).  The input resistance of 300~$\Omega$ was chosen to match
the wave impedance of the signal wire. A test pulse is applied to all
inputs via 3~k$\Omega$ resistors.

The TDC itself is built on two MTD133B integrated circuits \cite{mtd}
connected together by a common data bus. The MTD133B is an 8-channel
multihit time-to-digital converter, which measures the separation in
time between hits (transitions) arriving at the COMMON input and each
of eight independent channels. Each channel accepts up to 16 hits.
With the acquisition clock frequency of 250 MHz the time difference is
measured with the 0.5~ns least count in the 16-bit dynamic range.

In the acquisition mode the MTD133B continuously accepts the input hits
until the TDC board receives the STOP signal (ST line). Then the bus
driver disables the hit acquisition and starts data processing and
buffering. During this phase the MTD133B computes the time intervals
between the recorded hits and the common STOP signal and compares them
to the far time threshold. The data less than the threshold
are put on the common data bus. Note that after three ``pumping''
clocks a new valid data word is asserted on each readout clock until
all the recorded hits are processed.  The external magnitude
comparator (CMP) compares the time intervals to the near time
threshold. The data which exceed this threshold are loaded into the
buffer memory. Thus this buffer contains only hits which arrived at
the TDC within the preset time window. After the last recorded hit
has been processed, the TDC board sets a flag on the F line. The bus
driver waits until all TDCs in the segment complete data processing,
then switches the F lines to the buffer memory and reads out non-empty
memories in a priority order. This operation is completed when all TDC
boards reset the F flags.

The analog and digital parts of the printed circuit board have separate
ground and power planes which are connected to the power supply via the
differential LC filters. The discriminator outputs are also
differential. This technique provides an isolation factor more than
66~dB.

\bigskip
\begin{minipage}{\textwidth}
\begin{tabular}{ll}
\textbf{TDC board specification}\\[0.5ex]
Number of channels                  & 16 \\
Input impedance                     & 300 $\Omega$\\
Threshold range                     & 0.05--2 mA\\
Slewing  (from 2 to 20 thresholds)  & $<$1 ns\\
Maximum number of hits in each channel & 16\\
Double pulse resolution             & $<$10 ns\\
Least count                         & 0.5 ns\\
Time measurement range               & 20--2048 ns\\
Offset value                        & 0--2048 ns\\
Step of the range and offset setting & 8 ns\\
Time window width                   & 16--2048 ns\\
Step of the window setting,         & 8 ns\\
Differential nonlinearity           & $\pm$10\%\\
Time of data processing and readout & $\mathrm{(N_r + N_s +  3)\cdot100}$~ns 
\footnote{Here $\mathrm{N_r}$ is the number of hits recorded within the TDC time
range and $\mathrm{N_s}$ is the number of the selected hits.}
\\
Power dissipation                   & 4.3 Wt\\
\end{tabular}
\end{minipage}

\section{TDC testing}
\label{sec:test}

\begin{figure}[htbp]
\begin{center}
\epsfig{file=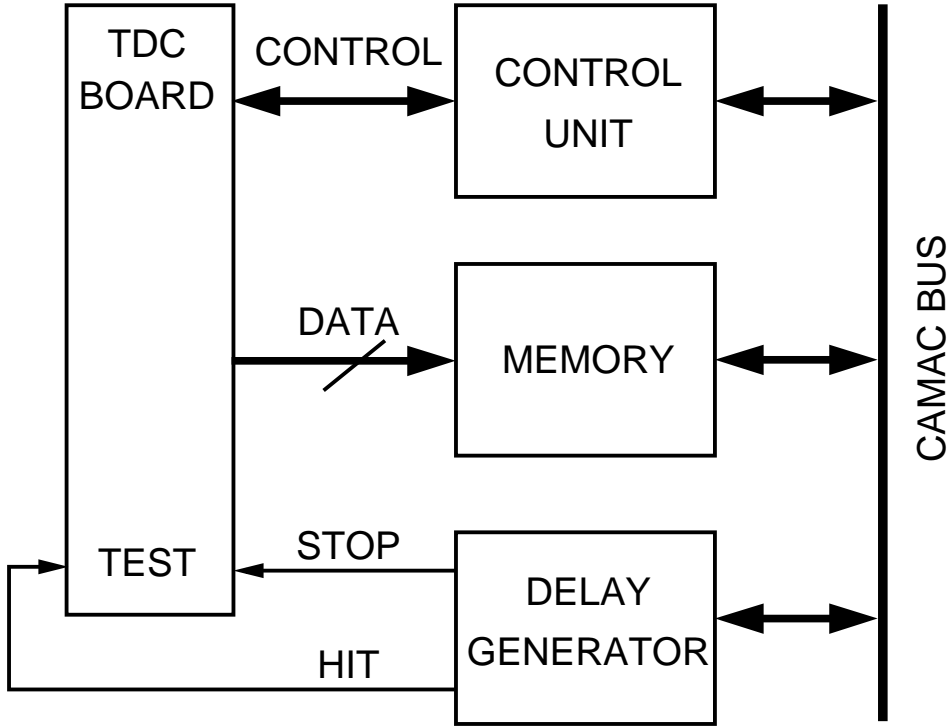,width=0.9\textwidth}
\end{center}
\caption{Test equipment.}
\label{fig3}
\end{figure}

The TDC board is designed to operate without tuning and therefore only
an after-production test is required. The test equipment
(Fig.\ref{fig3}) is based on the specially designed delay generator
operating in the range from 10~ns to 260~$\mu{}$s with the delay step of
20~ps. In the SINGLE HIT mode the module generates one hit and one STOP at
each trigger coming from the CAMAC bus.  In the BURST mode up to 16 hits and
one STOP are generated.

%
%
%

A set of programs has been written to test all components of the TDC
board. They allows to test and visualise the performance of each TDC
channel in the range of time interval between HIT and STOP from 20~ns
to 2048~ns with the step of 20~ps in the SINGLE HIT or BURST mode.
Also there is a possibility to scan automatically all positions of
the time window in each TDC channel to test the efficiency of the hit
selection.

\section{Conclusion}
\label{sec:conc}

The features of the above-described readout system result in the
reduced number of modules (there are no separate discriminator boards
and TDC modules), a small numbers of cables and high noise immunity.
At the threshold of 0.1~mA the drift chamber efficiency is about 99\%
and the spatial resolution equals to 85~$\mu$m \cite{dc}.

The readout system has been successfully operating in the DIRAC
experiment since 1999. At a typical rate about $6\cdot10^5$ counts per
spill (0.45~s) in each arm of the spectrometer the average access time
is about 2~$\mu$s. The accepted data are transferred to the VME
buffers, on average, within 5~$\mu{}$s.  This operation is not the
longest readout of the whole setup and thus it does not increase the
setup dead time.

\section*{Acknowledgments}

We would like to thank L.Nemenov and M.Ferro-Luzzi for support and
coordination of this work, V.Kruglov for fruitful discussions and
participation in the common job for the system implementation,
A.Kulikov for support of the setup trigger system, V.Olshevsky and
S.Trusov for readout software support and their help in testing the
TDC board prototype. The work was partially supported by the Russian
Foundation for Basic Research, project 01--02--17756.

\end{document}